\documentclass[authoryear,preprint,review,12pt]{elsarticle}

\usepackage{amssymb}
\usepackage{amsmath}

\usepackage{siunitx}
\DeclareSIUnit{\pixels}{pixels}

\journal{Journal of Food Engineering}

\begin{document}

\begin{frontmatter}



\title{3D evolution of protein networks and lipid globules in heat-treated egg yolk.} 


\author[a,b]{Felix Wittwer\corref{cor1}} 
\ead{felix.wittwer@desy.de}
\author[b]{Nimmi Das Anthuparambil} 
\author[b]{Frederik Unger} 
\author[b]{Randeer Pratap Gautam} 
\author[c]{Silja Flenner} 
\author[c]{Imke Greving}
\author[b]{Christian Gutt} 
\author[a,b]{Peter Modregger} 

\cortext[cor1]{Corresponding author}
\affiliation[a]{organization={Deutsches Elektronen-Synchrotron DESY},
                addressline={Notkestr. 85},
                postcode={22607},
                postcodesep={},
                city={Hamburg},
                country={Germany}
                }
\affiliation[b]{organization={Department Physik, Universität Siegen},
                addressline={Walter-Flex-Straße 3},
                postcode={57072},
                postcodesep={},
                city={Siegen},
                country={Germany}
                }
\affiliation[c]{organization={Helmholtz-Zentrum Hereon},
                addressline={Max-Planck-Straße 1},
                postcode={21502},
                postcodesep={},
                city={Geesthacht},
                country={Germany}
                }


\begin{abstract}
Upon heating, egg yolk transforms from a liquid to a gel due to protein denaturation.
This process can serve as a useful model to better understand protein denaturation in general.
Using x-ray holographic tomography, we investigated the structural changes in egg yolk during boiling, without the need for complex sample fixation or drying.
Our results reveal a developing separation between proteins and lipids, with fatty components rapidly aggregating into large globules that subsequently evolve into bubbles.
\end{abstract}


\end{frontmatter}



\section{Introduction}

Egg yolk is widely used as a versatile ingredient in cooking and food processing.
It serves as a natural emulsifier, binding fats and water to create smooth textures in sauces, dressings, and baked goods.
Beyond its culinary appeal, egg yolk is packed with essential nutrients, including proteins, lipids, vitamins, and minerals \citep{huopalahti2007bioactive}.
Cooking egg yolk involves a fascinating transformation of its proteins and lipids \citep{anthuparambil2023exploring}.
When heated, egg yolk undergoes denaturation, aggregation, coagulation and gelation, leading to changes in texture and consistency \citep{anthuparambil2023exploring,anton2013egg}.
The final texture depends on factors such as temperature and heating rate, demonstrating the delicate interplay of science and technique in cooking with egg yolk.
In general, the unique microstructure of a protein gel defines its viscoelasticity, with the resulting mechanical, interfacial, and transport properties being crucial for various applications in bio-nanotechnology and food industry \citep{mine2008egg}.


Although the liquid-to-gel transformation of egg yolk is a well-known phenomenon, the evolution of its microstructure during denaturation-aggregation-gelation processes is not yet fully understood.
In the past, electron microscopy (EM) \citep{cordobes2004rheology,aguilar2019heat} has been used to study the microstructure of gelatinous egg yolk.
However, chemical fixation of samples for this technique can lead to sample degradation and hence unreliable results.
Additionally, EM requires ultra-thin sectioning of samples due to the limited penetration depth of electrons while X rays can penetrate deeper into biological tissues and materials, providing a more comprehensive structural analysis while preserving the sample’s integrity, indicating the advantage of x-ray techniques. 


In previous works \citep{anthuparambil2023exploring,anthuparambil2024salt}, we used low-dose x-ray photon correlation spectroscopy (XPCS) to investigate the functional contribution of yolk components\textemdash plasma proteins, low-density lipoproteins (LDL), and egg yolk granules\textemdash to the formation of the grainy-gel microstructure under heat induction. We found that at temperatures above \SI{75}{\degreeCelsius}, two kinds of structure formation occur in egg yolk. Egg yolk plasma proteins form a gel network and LDLs form aggregates, which will be embedded in the protein gel network. However, the studies using small-angle and ultra-small-angle scattering could only provide insight into structure sizes smaller than \qty{600}{\nano\meter}, which is not large enough to resolve the $\gtrapprox$ \qty{1}{\micro\meter} features of the protein-LDL aggregate microstructure.

X-ray tomography allows to study the internal structure of objects without damaging or destructive sample preparation \citep{withersXrayComputedTomography2021}.
Conventional x-ray tomography measures the absorption contrast of the sample, which is negligible for soft, biological materials such as cooked egg yolk.
For these samples, x-ray phase contrast imaging has become the state of the art since the phase contrast can be hundreds of times larger than the absorption contrast \citep{paganinXrayPhasecontrastImaging2021}.
Phase contrast methods such as x-ray holography can measure small contrast differences with a minimal radiation dose to the sample \citep{flennerHardXrayNanoholotomography2020a}.
%
Using x-ray holography, we achieved a resolution of \qty{200}{\nano\meter}, which is complementary to our previous XPCS results and allows to relate the microtopology with the collective dynamics.
By measuring egg yolk samples heated for different time durations, we were able to image and follow the evolution of protein networks and lipid aggregation.

Boiling egg yolk not only denaturates proteins but also causes lipids to aggregate.
This aggregation has been previously studied with EM, requiring extensive sample preparation \citep{bellairsSTRUCTUREYOLKHENS1961, hsuHistologicalStructuresNative2009}.
X-ray tomography enabled us to study the lipids without fixing and potentially changing the sample.
Apart from the known polyhedral yolk spheres \citep{liuEffectYolkSpheres2023}, we also observed the formation of round lipid spheres with a singular round intrusion.

This article is organized as follows: In Section 2, we describe the experimental details, which encompasses the sample preparation, measurement and data reconstruction; In Section 3, we describe the observed evolution of the protein networks during heating; In Section 4, we report the formation of lipid globules, their transformation and provide a hypothesis for their evolution; Section 5 summarizes and discusses the results.

\section{Experimental details}
\subsection{Sample preparation}
The egg yolk used in this study was extracted from a hen egg purchased from a local supermarket. The extraction process was carried out as follows:
\begin{enumerate}
    \item The egg yolk was separated from the egg white using a steel strainer and washed with Milli-Q water to remove any residual albumen. 
    \item The cleaned yolk was placed on filter paper to absorb excess water and albumen. To ensure complete removal of albumen, the yolk was gently rolled on the filter paper several times.
    \item The vitelline membrane was punctured using a plastic pipette tip, and the yolk was extracted and stored in a 15 mL Falcon tube at \SI{5}{\degreeCelsius}.
\end{enumerate}
For tomography experiments, the yolk was filled into Kapton capillaries of three sizes, coined \emph{small}, \emph{intermediate}, and \emph{medium} sized.
The small capillaries have an inner diameter of \qty{0.12}{\milli\meter} and an outer diameter of \qty{0.17}{\milli\meter}. 
The intermediate capillaries have an inner diameter of \qty{0.25}{\milli\meter} and an outer diameter of \qty{0.3}{\milli\meter}.
The medium capillaries have an inner diameter of \qty{0.5}{\milli\meter} and an outer diameter of \qty{0.55}{\milli\meter}. Both ends of the capillaries were closed using epoxy glue. The sealed sample capillaries were dipped in boiling water at \SI{100}{\degreeCelsius} for heating times of up to \qty{22}{\minute}. After heating for a specific waiting time, samples were immediately quenched to low temperatures to stop further changes by dipping in ice water at $\approx$ \SI{5}{\degreeCelsius}. Afterwards, the sample capillaries were glued onto a vertical pin which was then mounted onto the sample holder for tomographic studies. 
The samples remained sealed in the capillaries with no further sample fixation necessary.

\subsection{Measurements}
\begin{figure}
    \begin{center}
        \includegraphics{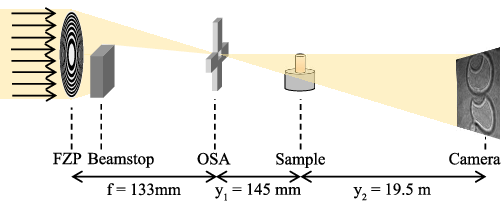}
    \end{center}
    \caption{Holo-tomography setup at the P05 beamline.}
    \label{fig:setup}
\end{figure}
We performed the measurements at the P05 beamline at Petra~III \citep{flennerHardXrayNanotomography2022a}, using the holo-tomography setup pictured in Fig.~\ref{fig:setup}.
The setup used a Si-111 double-crystal monochromator to filter the x-ray energy to \qty{11}{\kilo\electronvolt}; a \qty{300}{\micro\meter} diameter Fresnel zone plate with an outermost zone width of \qty{50}{\nano\meter}; a beamstop and an order-sorting aperture (OSA) to block unwanted diffraction from the zone plate.
The sample was installed on a rotation stage \qty{145.4}{\milli\meter} downstream of the focal plane.
After illuminating the sample, the X rays propagated through an evacuated pipe to the camera detector installed \qty{19.6}{\meter} downstream of the focal plane.

The holograms were recorded using a \qty{10}{\micro\meter}-thick Gadox scintillator that was directly coupled to a Hamamatsu C12849-101U sCMOS camera.
The camera had \numproduct{2048 x 2048}~\unit{\pixels} with a pixel size of \qty{6.5}{\micro\meter}.
For data processing, the pixels were \numproduct{2 x 2} binned, resulting in a pixel size of \qty{13}{\micro\meter} and an image size of \numproduct{1024 x 1024}~\unit{\pixels}.
This setup had a geometric magnification of \num{135} and achieved an effective pixel size of \qty{96}{\nano\meter}.
The effective field of view on the detector was \qty{98}{\micro\meter}.
Because this was too small to completely image the more than \qty{170}{\micro\meter}-wide sample capillary, we used local tomography to record only a \qty{98}{\micro\meter}-cylinder at the center of the capillaries.
For each tomogram, the sample was rotated continuously over \qty{180}{\degree} at \qty{0.125}{\degree} per second.
While the sample was rotated, the camera recorded holograms with an exposure time of \qty{0.8}{\second}.
A complete tomogram contains \num{1776}~holograms due to a small overhead between each exposure.
The radiation dose per scan was \qty{170}{\kilo\gray}.

%
%
%

\subsection{Data reconstruction}
Each hologram was automatically reconstructed on the Max\-well computing cluster using the Ho\-lo\-wi\-zard package to obtain phase projections of the sample \citep{doraArtifactsuppressingReconstructionStrongly2024, doraModelbasedAutofocusNearfield2025, doraPythonFrameworkOnline2024}.
From the phase projections, we reconstructed the 3D phase shift of the sample with a gridrec algorithm, a Shepp-Logan filter and Fourier-wavelet ring removal using the P05 reconstruction pipeline based on TomoPy \citep{gursoyTomoPyFrameworkAnalysis2014}.
The phase shift $\phi$ in each voxel is related to the refractive index decrement~$\delta$ via the wavenumber~$k$ and voxel size~$v$:
\begin{equation}
    \delta = - \frac{\phi}{k v}.
\end{equation}
Because the data was measured with local tomography, there is no zero reference to calibrate the phase shift offset.
Without knowing the absolute phase offset, we normalized the data by subtracting the average phase.
This is based on the assumption that the average phase shift is the same for every sample and the fact that the probed volume is much larger than the typical structure size of the network.
To suppress noise, a Gaussian filter with a standard deviation of $2v$ was applied to the reconstructed volumes.

\section{Evolution of protein networks}
\subsection{Network evolution}
\begin{figure}
    \begin{center}
        \includegraphics{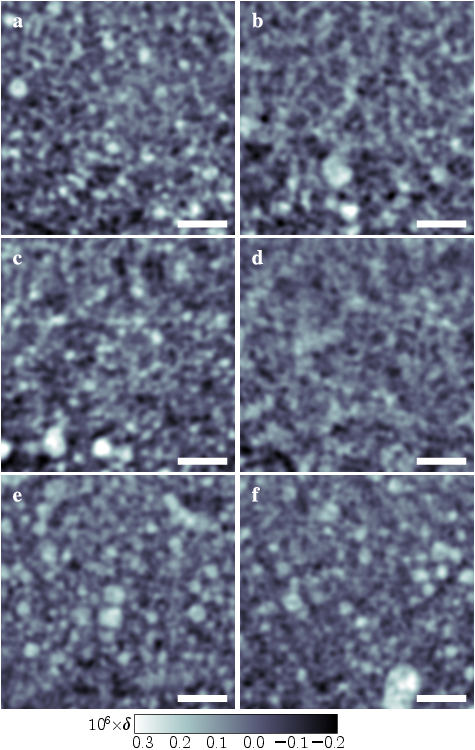}
    \end{center}
    \caption{Reconstructed slices of the refractive index decrement $\delta$. The samples were filled in small capillaries and heated to \qty{100}{\celsius} for (a) \qty{2}{\minute}, (b) \qty{3}{\minute}, (c) \qty{4}{\minute}, (d) \qty{5}{\minute}, (e) \qty{6}{\minute}, (f) \qty{7}{\minute}. For all images, the average value has been subtracted so that the images share the same color scale.
    The scale bars indicate \qty{5}{\micro\meter}.}
    \label{fig:network_overview}
\end{figure}
Raw egg yolk is difficult to measure with tomography because the liquid does not follow the rotation of the capillary.
Only after two minutes of heating was the egg yolk stiff enough to successfully record a tomogram.
We recorded the evolution of the protein networks during the first seven minutes of heating.
Figure.~\ref{fig:network_overview} shows example sections from different timesteps.
The images show the reconstructed refractive index decrement $\delta$, which is related to the refractive index $n$ via $n=1-\delta$.
In the images, lighter structures indicate larger $\delta$~values and correspond to regions of higher electron density.
In order of increasing electron density, egg yolk is mainly composed of fat, water, and protein.
Light regions therefore contain more proteins while dark regions contain more fats.

The samples already show a clear separation into protein-rich and fat-rich globules after \qty{2}{\minute} of heating, see Fig.~\ref{fig:network_overview}(a).
After this short period of heating, the protein structures in the yolk are small, most are less than one micrometer in size. 
Still, the protein globules are connected by a thin network of filaments that extends through the egg yolk.
After three minutes of heating, see Fig.~\ref{fig:network_overview}(b), protein and fat globules separate further and accumulate into larger structures.
The protein network becomes more pronounced and forms veined structures.
Further heating continues this trend as the scale of the network becomes larger and the protein- and fat-rich regions increase in size, see Fig.~\ref{fig:network_overview}(c-f).

\subsection{Structure analysis}
\begin{figure} %
    \begin{center}
        \includegraphics{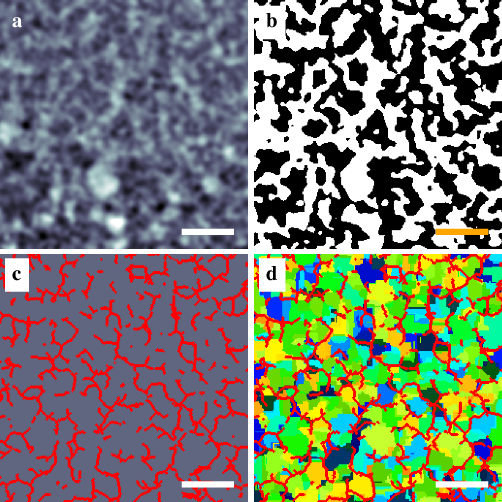}
    \end{center}
    \caption{(a) Example slice from the \qty{3}{\minute} sample. (b) Thresholded slice, white areas are greater than the average, black areas smaller. (c) Extracted network combining ten neighboring slices to the example slice. (d) Segmented voids between the network. The scale bars indicate \qty{5}{\micro\meter}.}
    \label{fig:network_extraction}
\end{figure}

To analyze the morphology of the protein networks, we extracted the network structure from the reconstructed volumes.
To avoid potential edge artifacts from local tomography, we extracted a \numproduct{456 x 456 x 456}~voxel subvolume from each dataset.
Next, we subtracted an estimated background from the data to improve the correct classification of the network.
Because the feature size of the network was not larger than a few micrometer, we estimated the background by filtering the data with a $10v$ Gaussian filter.
With the background removed, we converted the data to a binary mask by setting all positive voxels to one and the rest to zero, see Fig.~\ref{fig:network_extraction}(b).
From the binary mask, we extracted the structure of the network with the skeletonize function from scikit-image \citep{vanderwaltScikitimageImageProcessing2014}, shown in Fig.~\ref{fig:network_extraction}(c).
To analyze the structure size, we segmented the voids in the network using the watershed function from scikit-image, see Fig.~\ref{fig:network_extraction}(d).
On average, the size of the segments is around \qty{880}{\nano\meter}.
We also recorded a second time series in intermediate-sized capillaries. 
For these samples, the average segment size varies remarkably between \qtyrange{780}{880}{\nano\meter}, much further than in the first time series.

\section{Evolution of lipid globules}
\subsection{Globule formation}
Because the sample preparation was much quicker than the time it took to collect one tomogram, the prepared samples had to be stored until they could be measured.
The samples were stored at room temperature, sometimes for more than an hour. 
During this time, some of the samples unexpectedly developed large lipid globules.

\begin{figure} %
    \begin{center}
        \centerline{ 
            \includegraphics{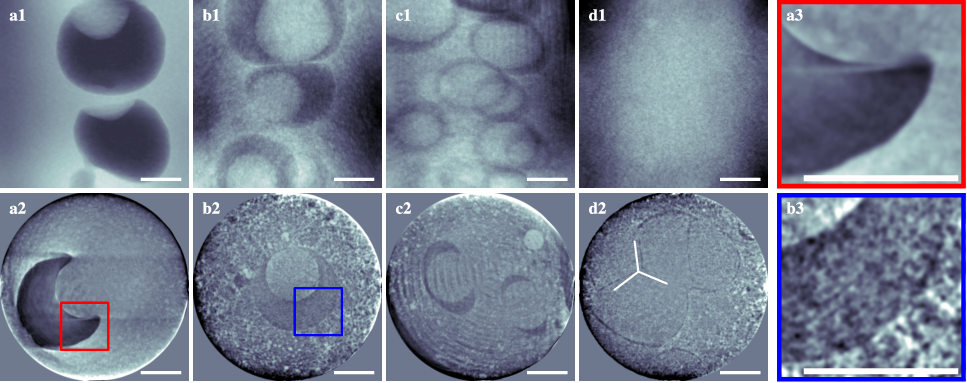}
            }
    \end{center}
    \caption{Lipid globules arranged according to the estimated evolution. (a1-d1) show projections of the samples. (a2-d2) show horizontal slices through the sample. The horizontal stripes in (a2) are reconstruction artifacts caused by small sample changes during the measurement. (d2) contains a marker indicating the \qty{120}{\degree} intersection angles. (a3-b3) show enlarged sections for the first two samples. All scale bars indicate \qty{20}{\micro\meter}. The color scale of each image is individually optimized for best contrast.}
    \label{fig:lipid_bubbles}
\end{figure}
Figure~\ref{fig:lipid_bubbles} shows four exemplary scans, arranged according to our estimated evolution of the lipid globules. 
In the first three columns, the egg yolk was inside small capillaries and heated for \qty{6}{\minute}, \qty{1.5}{\minute}, and \qty{4}{\minute}, respectively.
The sample shown in Fig.~\ref{fig:lipid_bubbles}d was in a medium capillary and heated for \qty{22}{\minute}.

Heating egg yolk not only denaturates the proteins but also affects the phospholipids and lipoproteins.
At higher temperatures, the phospholipid membranes become more permeable and can release the lipids inside.
According to our estimate, the lipids agglomerate and form large globules.
It is peculiar that these globules have exactly one intrusion, see Fig.~\ref{fig:lipid_bubbles}(a).
This intrusion grows and the globules begin to become hollow, see Fig.~\ref{fig:lipid_bubbles}(b).
Notably, the intrusions are highly spherical, independent of the size and outer shape of the globule.
This hollowing trend continues until the intrusions fill out most of the globule, see Fig.~\ref{fig:lipid_bubbles}(c).
Remarkably, the intrusions are still spherical at this stage, although the outer shape has become prolate.
The circular shape of the intrusion points to a strong surface tension at the interface.
At the end of the globule evolution, the lipid shell becomes thin and bubble-like, see Fig.~\ref{fig:lipid_bubbles}(d).
These thin structures appear only weakly in the projection image but are readily apparent in the slice image.
As the globules grow, they probably come into contact with neighboring globules but do not seem to merge together.
Instead, they remain separated by a wall that is straight or has only a low curvature.
When three walls meet, the angles between the walls are near \qty{120}{\degree}, exemplary indicated by the marker in Fig.~\ref{fig:lipid_bubbles}(d2).
These angles are congruent with Plateau's laws for bubble surfaces at equilibrium \citep{ballShapes2011} and also indicative of a strong surface tension.
Figure~\ref{fig:lipid_bubbles}(a3) and (b3) show enlarged sections of lipid globules and highlight that the protein network permeates through some of the lipid globules.

\subsection{Hypothesis of globule evolution}
\begin{figure} %
    \begin{center}
        \centerline{ 
            \includegraphics{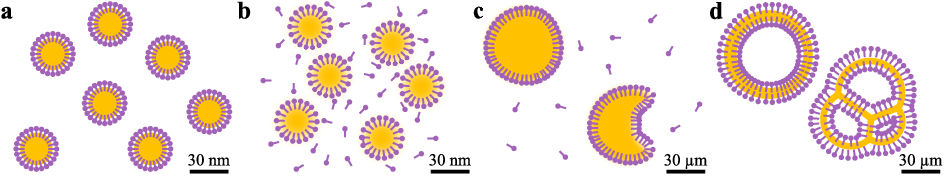}
            }
    \end{center}
    \caption{Hypothesized evolution of the lipid globules. (a) Lipoproteins in raw egg yolk. (b) At high temperatures, the phospholipid membrane becomes permeable and the fat leaks out. (c) Large fat droplets have a smaller surface-to-volume ratio, inducing shape changes. (d) Ultimately, the globules become hollow shells, similar to soap bubbles. }
    \label{fig:bubble_formation}
\end{figure}
The lipoproteins in raw egg yolk are about \qty{30}{\nano\meter} in size \citep{evansEggYolkVery1974} with a surface-to-volume ratio (SA:V) of \qty{0.1}{\per\nano\meter}.
When the egg yolk is heated, the phospholipid membrane becomes more permeable and releases its fat content, pictured in Fig.~\ref{fig:bubble_formation}(b).
The free-floating fat merges into larger fat bubbles with a much smaller SA:V.
Because phospholipids are amphiphilic, they aggregate at the water-oil-interface.
This creates a pressure on the shape of the globules to increase the SA:V, see Fig.~\ref{fig:bubble_formation}(c).
Note the scalebar indicating that these globules are much larger than the original lipoproteins.
Ultimately, to maximize the surface area, the globules become shells, illustrated in Fig.~\ref{fig:bubble_formation}(d).
Without other influences, the SA:V of the shell at the end of the globule evolution would be the same as in the original lipoproteins.

\section{Discussion}
We successfully imaged protein networks to study their evolution in heat-treated egg yolk.
In the first five minutes of heating, the network grows and its structure becomes coarser.
Simultaneously, the protein- and fat-rich regions become separate and more distinct.
While the evolution for one time series was clear, a second time series showed different quantitative values, indicating the need for more measurements to improve the statistics.

The fat in the egg yolk also changes from lipoproteins to lipid globules.
These globules become hollow and ultimately transform into bubbles with a size of \qtyrange{50}{100}{\micro\meter}.
This behaviour is probably caused by the dissolution of lipoproteins at high temperatures, leading to an out-flow of fat that accumulates into globules which than transform into bubbles.
Fully understanding this effect requires further investigations.

\section{Acknowledgements}
This research was supported in part through the Maxwell computational resources operated at Deutsches Elektronen-Synchrotron DESY, Hamburg, Germany.
We acknowledge DESY (Hamburg, Germany), a member of the Helmholtz Association HGF, for the provision of experimental facilities.
Parts of this research were carried out at PETRA III and we would like to thank the staff for assistance in using the P05 beamline.
Beamtime was allocated for proposal I-20240551.

\section{Data availability}
Data underlying the results presented in this paper are not publicly available at this time but may be obtained from the authors upon reasonable request.

\section{CRediT authorship contribution statement}
to be completed





\bibliographystyle{elsarticle-harv} 
\bibliography{iucr}

\begin{thebibliography}{21}
\expandafter\ifx\csname natexlab\endcsname\relax\def\natexlab#1{#1}\fi
\providecommand{\url}[1]{\texttt{#1}}
\providecommand{\href}[2]{#2}
\providecommand{\path}[1]{#1}
\providecommand{\DOIprefix}{doi:}
\providecommand{\ArXivprefix}{arXiv:}
\providecommand{\URLprefix}{URL: }
\providecommand{\Pubmedprefix}{pmid:}
\providecommand{\doi}[1]{\href{http://dx.doi.org/#1}{\path{#1}}}
\providecommand{\Pubmed}[1]{\href{pmid:#1}{\path{#1}}}
\providecommand{\bibinfo}[2]{#2}
\ifx\xfnm\relax \def\xfnm[#1]{\unskip,\space#1}\fi
\bibitem[{Aguilar et~al.(2019)Aguilar, Cordob{\'e}s, Bengoechea and
  Guerrero}]{aguilar2019heat}
\bibinfo{author}{Aguilar, J.}, \bibinfo{author}{Cordob{\'e}s, F.},
  \bibinfo{author}{Bengoechea, C.}, \bibinfo{author}{Guerrero, A.},
  \bibinfo{year}{2019}.
\newblock \bibinfo{title}{Heat-induced gelation of egg yolk as a function of
  p{H}. does the type of acid make any difference?}
\newblock \bibinfo{journal}{Food Hydrocolloids} \bibinfo{volume}{87},
  \bibinfo{pages}{142--148}.
\bibitem[{Anthuparambil et~al.(2023)Anthuparambil, Girelli, Timmermann,
  Kowalski, Akhundzadeh, Retzbach, Senft, Dargasz, Gutm{\"u}ller, Hiremath
  et~al.}]{anthuparambil2023exploring}
\bibinfo{author}{Anthuparambil, N.D.}, \bibinfo{author}{Girelli, A.},
  \bibinfo{author}{Timmermann, S.}, \bibinfo{author}{Kowalski, M.},
  \bibinfo{author}{Akhundzadeh, M.S.}, \bibinfo{author}{Retzbach, S.},
  \bibinfo{author}{Senft, M.D.}, \bibinfo{author}{Dargasz, M.},
  \bibinfo{author}{Gutm{\"u}ller, D.}, \bibinfo{author}{Hiremath, A.}, et~al.,
  \bibinfo{year}{2023}.
\newblock \bibinfo{title}{Exploring non-equilibrium processes and
  spatio-temporal scaling laws in heated egg yolk using coherent x-rays}.
\newblock \bibinfo{journal}{Nature Communications} \bibinfo{volume}{14},
  \bibinfo{pages}{5580}.
\bibitem[{Anthuparambil et~al.(2024)Anthuparambil, Timmermann, Dargasz,
  Retzbach, Senft, Begam, Ragulskaya, Paulus, Zhang, Westermeier
  et~al.}]{anthuparambil2024salt}
\bibinfo{author}{Anthuparambil, N.D.}, \bibinfo{author}{Timmermann, S.},
  \bibinfo{author}{Dargasz, M.}, \bibinfo{author}{Retzbach, S.},
  \bibinfo{author}{Senft, M.D.}, \bibinfo{author}{Begam, N.},
  \bibinfo{author}{Ragulskaya, A.}, \bibinfo{author}{Paulus, M.},
  \bibinfo{author}{Zhang, F.}, \bibinfo{author}{Westermeier, F.}, et~al.,
  \bibinfo{year}{2024}.
\newblock \bibinfo{title}{Salt induced slowdown of kinetics and dynamics during
  thermal gelation of egg-yolk}.
\newblock \bibinfo{journal}{The Journal of Chemical Physics}
  \bibinfo{volume}{161}.
\bibitem[{Anton(2013)}]{anton2013egg}
\bibinfo{author}{Anton, M.}, \bibinfo{year}{2013}.
\newblock \bibinfo{title}{Egg yolk: structures, functionalities and processes}.
\newblock \bibinfo{journal}{Journal of the Science of Food and Agriculture}
  \bibinfo{volume}{93}, \bibinfo{pages}{2871--2880}.
\bibitem[{Ball(2011)}]{ballShapes2011}
\bibinfo{author}{Ball, P.}, \bibinfo{year}{2011}.
\newblock \bibinfo{title}{Shapes}.
\newblock Number~\bibinfo{number}{1} in \bibinfo{series}{Nature's
  {{Patterns}}}. \bibinfo{edition}{paperback} ed., \bibinfo{publisher}{Oxford
  Univ. Press}, \bibinfo{address}{Oxford}.
\bibitem[{Bellairs(1961)}]{bellairsSTRUCTUREYOLKHENS1961}
\bibinfo{author}{Bellairs, R.}, \bibinfo{year}{1961}.
\newblock \bibinfo{title}{{{THE STRUCTURE OF THE YOLK OF THE HEN}}'{{S EGG AS
  STUDIED BY ELECTRON MICROSCOPY}}}.
\newblock \bibinfo{journal}{The Journal of Cell Biology} \bibinfo{volume}{11},
  \bibinfo{pages}{207--225}.
\newblock \DOIprefix\doi{10.1083/jcb.11.1.207}.
\bibitem[{Cordob{\'e}s et~al.(2004)Cordob{\'e}s, Partal and
  Guerrero}]{cordobes2004rheology}
\bibinfo{author}{Cordob{\'e}s, F.}, \bibinfo{author}{Partal, P.},
  \bibinfo{author}{Guerrero, A.}, \bibinfo{year}{2004}.
\newblock \bibinfo{title}{Rheology and microstructure of heat-induced egg yolk
  gels}.
\newblock \bibinfo{journal}{Rheologica Acta} \bibinfo{volume}{43},
  \bibinfo{pages}{184--195}.
\bibitem[{Dora et~al.(2024a)Dora, Flenner, Lopes~Marinho and
  Hagemann}]{doraPythonFrameworkOnline2024}
\bibinfo{author}{Dora, J.}, \bibinfo{author}{Flenner, S.},
  \bibinfo{author}{Lopes~Marinho, A.}, \bibinfo{author}{Hagemann, J.},
  \bibinfo{year}{2024}a.
\newblock \bibinfo{title}{A {{Python}} framework for the online reconstruction
  of {{X-ray}} near-field holography data}.
\newblock \bibinfo{howpublished}{Zenodo}.
\newblock \DOIprefix\doi{10.5281/ZENODO.8349364}.
\bibitem[{Dora et~al.(2025)Dora, M{\"o}ddel, Flenner, Reimers,
  {Zeller-Plumhoff}, Schroer, Knopp and
  Hagemann}]{doraModelbasedAutofocusNearfield2025}
\bibinfo{author}{Dora, J.}, \bibinfo{author}{M{\"o}ddel, M.},
  \bibinfo{author}{Flenner, S.}, \bibinfo{author}{Reimers, J.},
  \bibinfo{author}{{Zeller-Plumhoff}, B.}, \bibinfo{author}{Schroer, C.G.},
  \bibinfo{author}{Knopp, T.}, \bibinfo{author}{Hagemann, J.},
  \bibinfo{year}{2025}.
\newblock \bibinfo{title}{Model-based autofocus for near-field phase
  retrieval}.
\newblock \bibinfo{journal}{Opt. Express} \bibinfo{volume}{33},
  \bibinfo{pages}{6641}.
\newblock \DOIprefix\doi{10.1364/OE.544573}.
\bibitem[{Dora et~al.(2024b)Dora, M{\"o}ddel, Flenner, Schroer, Knopp and
  Hagemann}]{doraArtifactsuppressingReconstructionStrongly2024}
\bibinfo{author}{Dora, J.}, \bibinfo{author}{M{\"o}ddel, M.},
  \bibinfo{author}{Flenner, S.}, \bibinfo{author}{Schroer, C.G.},
  \bibinfo{author}{Knopp, T.}, \bibinfo{author}{Hagemann, J.},
  \bibinfo{year}{2024}b.
\newblock \bibinfo{title}{Artifact-suppressing reconstruction of strongly
  interacting objects in {{X-ray}} near-field holography without a spatial
  support constraint}.
\newblock \bibinfo{journal}{Opt. Express} \bibinfo{volume}{32},
  \bibinfo{pages}{10801}.
\newblock \DOIprefix\doi{10.1364/OE.514641}.
\bibitem[{Evans et~al.(1974)Evans, Bauer and Flegal}]{evansEggYolkVery1974}
\bibinfo{author}{Evans, R.J.}, \bibinfo{author}{Bauer, D.H.},
  \bibinfo{author}{Flegal, C.J.}, \bibinfo{year}{1974}.
\newblock \bibinfo{title}{The {{Egg Yolk Very Low Density Lipoproteins}} of
  {{Fresh}} and {{Stored Shell Eggs}}}.
\newblock \bibinfo{journal}{Poultry Science} \bibinfo{volume}{53},
  \bibinfo{pages}{645--652}.
\newblock \DOIprefix\doi{10.3382/ps.0530645}.
\bibitem[{Flenner et~al.(2022)Flenner, Hagemann, Storm, Kubec, Qi, David,
  Longo, Niese, Gawlitza, {Zeller-Plumhoff}, Reimers, M{\"u}ller and
  Greving}]{flennerHardXrayNanotomography2022a}
\bibinfo{author}{Flenner, S.}, \bibinfo{author}{Hagemann, J.},
  \bibinfo{author}{Storm, M.}, \bibinfo{author}{Kubec, A.},
  \bibinfo{author}{Qi, P.}, \bibinfo{author}{David, C.},
  \bibinfo{author}{Longo, E.}, \bibinfo{author}{Niese, S.},
  \bibinfo{author}{Gawlitza, P.}, \bibinfo{author}{{Zeller-Plumhoff}, B.},
  \bibinfo{author}{Reimers, J.}, \bibinfo{author}{M{\"u}ller, M.},
  \bibinfo{author}{Greving, I.}, \bibinfo{year}{2022}.
\newblock \bibinfo{title}{Hard x-ray nanotomography at the {{P05}} imaging
  beamline at {{PETRA III}}}, in: \bibinfo{editor}{M{\"u}ller, B.},
  \bibinfo{editor}{Wang, G.} (Eds.), \bibinfo{booktitle}{Developments in
  {{X-Ray Tomography XIV}}}, \bibinfo{publisher}{SPIE}, \bibinfo{address}{San
  Diego, United States}. p.~\bibinfo{pages}{19}.
\newblock \DOIprefix\doi{10.1117/12.2632706}.
\bibitem[{Flenner et~al.(2020)Flenner, Kubec, David, Storm, Schaber, Vollrath,
  M{\"u}ller, Greving and Hagemann}]{flennerHardXrayNanoholotomography2020a}
\bibinfo{author}{Flenner, S.}, \bibinfo{author}{Kubec, A.},
  \bibinfo{author}{David, C.}, \bibinfo{author}{Storm, M.},
  \bibinfo{author}{Schaber, C.F.}, \bibinfo{author}{Vollrath, F.},
  \bibinfo{author}{M{\"u}ller, M.}, \bibinfo{author}{Greving, I.},
  \bibinfo{author}{Hagemann, J.}, \bibinfo{year}{2020}.
\newblock \bibinfo{title}{Hard {{X-ray}} nano-holotomography with a {{Fresnel}}
  zone plate}.
\newblock \bibinfo{journal}{Opt. Express} \bibinfo{volume}{28},
  \bibinfo{pages}{37514}.
\newblock \DOIprefix\doi{10.1364/OE.406074}.
\bibitem[{G{\"u}rsoy et~al.(2014)G{\"u}rsoy, De~Carlo, Xiao and
  Jacobsen}]{gursoyTomoPyFrameworkAnalysis2014}
\bibinfo{author}{G{\"u}rsoy, D.}, \bibinfo{author}{De~Carlo, F.},
  \bibinfo{author}{Xiao, X.}, \bibinfo{author}{Jacobsen, C.},
  \bibinfo{year}{2014}.
\newblock \bibinfo{title}{{{TomoPy}}: A framework for the analysis of
  synchrotron tomographic data}.
\newblock \bibinfo{journal}{J Synchrotron Rad} \bibinfo{volume}{21},
  \bibinfo{pages}{1188--1193}.
\newblock \DOIprefix\doi{10.1107/S1600577514013939}.
\bibitem[{Hsu et~al.(2009)Hsu, Chung and
  Lai}]{hsuHistologicalStructuresNative2009}
\bibinfo{author}{Hsu, K.C.}, \bibinfo{author}{Chung, W.H.},
  \bibinfo{author}{Lai, K.M.}, \bibinfo{year}{2009}.
\newblock \bibinfo{title}{Histological {{Structures}} of {{Native}} and
  {{Cooked Yolks}} from {{Duck Egg Observed}} by {{SEM}} and {{Cryo-SEM}}}.
\newblock \bibinfo{journal}{J. Agric. Food Chem.} \bibinfo{volume}{57},
  \bibinfo{pages}{4218--4223}.
\newblock \DOIprefix\doi{10.1021/jf900495n}.
\bibitem[{Huopalahti et~al.(2007)Huopalahti, Anton, L{\'o}pez-Fandi{\~n}o and
  Schade}]{huopalahti2007bioactive}
\bibinfo{author}{Huopalahti, R.}, \bibinfo{author}{Anton, M.},
  \bibinfo{author}{L{\'o}pez-Fandi{\~n}o, R.}, \bibinfo{author}{Schade, R.},
  \bibinfo{year}{2007}.
\newblock \bibinfo{title}{Bioactive Egg Compounds}. volume~\bibinfo{volume}{5}.
\newblock \bibinfo{publisher}{Springer}, \bibinfo{address}{Berlin}.
\bibitem[{Liu et~al.(2023)Liu, Wang, Ma, Wang, Zhu and
  Jin}]{liuEffectYolkSpheres2023}
\bibinfo{author}{Liu, Y.}, \bibinfo{author}{Wang, K.}, \bibinfo{author}{Ma,
  J.}, \bibinfo{author}{Wang, Z.}, \bibinfo{author}{Zhu, Q.},
  \bibinfo{author}{Jin, Y.}, \bibinfo{year}{2023}.
\newblock \bibinfo{title}{Effect of yolk spheres as a key histological
  structure on the morphology, character, and oral sensation of boiled egg yolk
  gel}.
\newblock \bibinfo{journal}{Food Chemistry} \bibinfo{volume}{424},
  \bibinfo{pages}{136380}.
\newblock \DOIprefix\doi{10.1016/j.foodchem.2023.136380}.
\bibitem[{Mine(2008)}]{mine2008egg}
\bibinfo{author}{Mine, Y.}, \bibinfo{year}{2008}.
\newblock \bibinfo{title}{Egg bioscience and biotechnology}.
\newblock \bibinfo{publisher}{John Wiley \& Sons}.
\bibitem[{Paganin and Pelliccia(2021)}]{paganinXrayPhasecontrastImaging2021}
\bibinfo{author}{Paganin, D.M.}, \bibinfo{author}{Pelliccia, D.},
  \bibinfo{year}{2021}.
\newblock \bibinfo{title}{X-ray phase-contrast imaging: A broad overview of
  some fundamentals}, in: \bibinfo{booktitle}{Advances in {{Imaging}} and
  {{Electron Physics}}}. \bibinfo{publisher}{Elsevier}. volume
  \bibinfo{volume}{218}, pp. \bibinfo{pages}{63--158}.
\newblock \DOIprefix\doi{10.1016/bs.aiep.2021.04.002}.
\bibitem[{{van der Walt} et~al.(2014){van der Walt}, Sch{\"o}nberger,
  {Nunez-Iglesias}, Boulogne, Warner, Yager, Gouillart and
  Yu}]{vanderwaltScikitimageImageProcessing2014}
\bibinfo{author}{{van der Walt}, S.}, \bibinfo{author}{Sch{\"o}nberger, J.L.},
  \bibinfo{author}{{Nunez-Iglesias}, J.}, \bibinfo{author}{Boulogne, F.},
  \bibinfo{author}{Warner, J.D.}, \bibinfo{author}{Yager, N.},
  \bibinfo{author}{Gouillart, E.}, \bibinfo{author}{Yu, T.},
  \bibinfo{year}{2014}.
\newblock \bibinfo{title}{Scikit-image: Image processing in {{Python}}}.
\newblock \bibinfo{journal}{PeerJ} \bibinfo{volume}{2}, \bibinfo{pages}{e453}.
\newblock \DOIprefix\doi{10.7717/peerj.453}.
\bibitem[{Withers et~al.(2021)Withers, Bouman, Carmignato, Cnudde, Grimaldi,
  Hagen, Maire, Manley, Du~Plessis and
  Stock}]{withersXrayComputedTomography2021}
\bibinfo{author}{Withers, P.J.}, \bibinfo{author}{Bouman, C.},
  \bibinfo{author}{Carmignato, S.}, \bibinfo{author}{Cnudde, V.},
  \bibinfo{author}{Grimaldi, D.}, \bibinfo{author}{Hagen, C.K.},
  \bibinfo{author}{Maire, E.}, \bibinfo{author}{Manley, M.},
  \bibinfo{author}{Du~Plessis, A.}, \bibinfo{author}{Stock, S.R.},
  \bibinfo{year}{2021}.
\newblock \bibinfo{title}{X-ray computed tomography}.
\newblock \bibinfo{journal}{Nat Rev. Methods Primers} \bibinfo{volume}{1},
  \bibinfo{pages}{18}.
\newblock \DOIprefix\doi{10.1038/s43586-021-00015-4}.

\end{thebibliography}
\end{document}